\documentstyle[preprint,eqsecnum,aps]{revtex}

\begin{document}
\draft
\preprint{}
\title{The connection between superconducting phase correlations and spin excitations in 
YBa$_2$Cu$_3$O$_{6.6}$: A magnetic field study  }
\author{Pengcheng Dai$^\ast$, H. A. Mook$^*$, G. Aeppli$^\dag$, S. M. Hayden$^\ddag$  
 \& F. Do$\rm\breve{g}$an$^\S$}
\address{
$^\ast$Oak Ridge National Laboratory, Oak Ridge, 
Tennessee 37831-6393, USA\\
$^\dag$NEC Research Institute, Princeton, New Jersey 08540, USA\\
$^\ddag$H. H. Wills Physics Laboratory, University of Bristol, Bristol BS8 ITL, UK \\
$^\S$Department of Materials Science and Engineering, 
University of Washington, Seattle, Washington 98195, USA\\
}
\maketitle
\narrowtext
{\bf One of the most striking universal properties of the high-transition-temperature (high-$T_c$) superconductors is that they are all derived from the hole-doping of their insulating antiferromagnetic (AF) parent compounds. From the outset, the intimate relationship between magnetism and superconductivity in these copper-oxides has intrigued researchers \cite{anderson,emery,hirsch,schulz}. Evidence for this link comes from neutron scattering experiments that show the unambiguous presence of short-range AF correlations (excitations) in cuprate superconductors. Even so, the role of such excitations in the pairing mechanism and superconductivity is still a subject of controversy \cite{scalapino}. For YBa$_2$Cu$_3$O$_{6+x}$, where $x$ controls the hole-doping level, the most prominent feature in the magnetic excitations spectra is the 
``resonance'' \cite{mignod,mook,fong,dai1,fong1,dais}. Here we show that for underdoped YBa$_2$Cu$_3$O$_{6.6}$, where $x$ and $T_c$ are below the optimal values, modest magnetic fields suppress the resonance significantly, much more so for fields approximately perpendicular rather than parallel to the CuO$_2$ planes. Our results indicate that the resonance measures pairing and phase coherence, suggesting that magnetism plays an important role in the superconductivity of cuprates. The persistence of a field effect above $T_c$ favors mechanisms with preformed pairs in the normal state of underdoped cuprates \cite{uemura,emery1,lee}.
}

Neutrons carry magnetic dipoles which are scattered by the electron spins in the sample with a probability, dependent on the neutron direction and energy, directly proportional to the Fourier transform in space and time of the two-spin correlation function, $S(Q,\omega)$. The wave-vector $Q$ and energy $\hbar\omega$ are simply the difference between the momenta and energies, respectively, of the incident and scattered neutrons. YBa$_2$Cu$_3$O$_{6+x}$ [denoted as (123)O$_{6+x}$] with $x$ of 0.6 was selected because it is a thoroughly characterized high-$T_c$ superconductor of which it is possible to fabricate the large single crystals required for neutron measurements. Equally important, thermodynamic data indicate that even modest fields have a large effect on the electronic entropy in the superconducting phase \cite{junod1,junod2}. Thus, if the electronic entropy changes around $T_c$ 
are mostly due to spin excitations \cite{dais}, they must respond to an external magnetic field in qualitatively the same way \cite{janko}.

(123)O$_{6+x}$ contains pairs of CuO$_2$ layers separated by charge reservoir layers. 
We label wave vectors $Q = (q_x, q_y, q_z)$ in units of \AA\ as 
$(H, K, L) = (q_xa/2\pi, q_yb/2\pi, q_zc/2\pi)$ in the reciprocal lattice units (rlu) 
appropriate for the orthorhombic unit cell, for which the lattice parameters are $a = 3.83$, 
$b = 3.88$ ($\approx a$), and $c = 11.743$ \AA\ if $x = 0.6$. For $x = 0$, (123)O$_{6+x}$ 
is an antiferromagnetic (AF) insulator where the spin on each Cu$^{2+}$ ion is antiparallel 
to its nearest neighbors both within as well as in the immediately adjacent layer \cite{tranquada}. 
As the oxygen content $x$ is increased, the material becomes a metal and eventually, near $x = 1$, 
an excellent high-temperature ($T_c \sim 93$ K) superconductor. At $x$ near 0.6, (123)O$_{6+x}$ is also a good superconductor with $T_c$ reduced to $\sim$60 K. During the evolution from insulator to superconductor, static antiferromagnetism is lost while magnetic (spin) fluctuations centered at the $x = 0$ AF Bragg positions, often referred to as $Q_0 = (\pi,\pi)$ (refs \cite{mignod,mook,fong,dai1,fong1,dais}), persist. 
For highly doped (123)O$_{6+x}$, the most prominent feature in the spin fluctuation spectrum is a 
sharp resonance which appears below $T_c$ at an energy of 41 meV (refs \cite{mignod,mook,fong}). 
When scanned at fixed frequency as a function of wave vector, the sharp peak is centered at $(\pi,\pi)$ 
(refs \cite{mignod,mook,fong}) and its intensity is unaffected by a 11.5 T field in the ab-plane \cite{bourgesf}. 
In our underdoped (123)O$_{6.6}$ ($T_c = 62.7$ K) (ref. \cite{dai1}),  
the resonance occurs at 34 meV and is superposed on a continuum which is gapped at low energies \cite{dais}.  
For frequencies below the resonance, the continuum is actually strongest at a quartet of incommensurate positions disposed symmetrically about $(\pi,\pi)$ (refs \cite{dai2,mook1}). Both the resonance and the incommensurate fluctuations appear in the acoustic channel for the bilayers, meaning that they are due to spin correlations which are AF between the planes. The corresponding neutron scattering intensities exhibit a modulation $\sin^2(\pi dL/c)$, 
where $d$ (3.342 \AA) is the spacing between the nearest-neighbor CuO$_2$ planes, along $c$. 
In contrast, optical fluctuations are associated with ferromagnetic correlations between 
planes and vary as $\cos^2(\pi dL/c)$ along $c$. The optical fluctuations for $x = 0.6$ have a gap of $\sim$50 meV, well above the acoustic gap of 20 meV (ref. \cite{dais}).

We mounted the crystal in three different orientations inside a split coil 
(to provide neutron beam access) 7 T vertical field superconducting magnet to establish the effects of a field approximately along the c-axis ($B\sim||c$-axis, Fig. 1a) on the acoustic, 
along the $c$-axis ($B||c$-axis) on the optical, and in the ab-plane ($B||ab$-plane, Fig. 1d) on the acoustic fluctuations. Raw neutron scattering data for (123)O$_{6+x}$ in the large frequency range covered here contain considerable non-magnetic background. The background, however, is weakly temperature-dependent for $T < 70$ K, 
which makes temperature difference spectra a useful guide to the magnetic contributions. 
Figure 1b shows the effect of superconductivity on $S(Q,\omega)$ at $Q = (1/2, 3/2, 1.8)$ in the form of difference spectra between 10 K ($< T_c$) and 70 K ($T_c + 7.3$ K). At zero field ($B = 0$ T), 
we reproduce previous results \cite{dai1} in both the $B\sim||c$-axis (Fig. 1b) and the $B||ab$-plane 
(Fig. 1e) sample orientations. The data reveal the positive peak at 34 meV due to the increased resonance amplitude below $T_c$. Imposition of a 6.8 T field nearly along the $c$-axis results in the qualitatively similar temperature difference spectrum (Fig. 1c). However, the intensity gain of the resonance is suppressed by $\sim$30\% compared to that at zero field (Fig. 1b). This is remarkable because the magnitude (6.8 T) of the applied field is much less than the upper critical field ($B_{c2} = 45$ T) for the material \cite{junod2}, and the corresponding Zeeman (magnetic) energy is, at $\pm g\mu_B$ ($\sim0.8$ meV assuming $g = 2$ and $S = 1$), much smaller than the 34 meV resonance energy and the thermal energy $k_BT_c$ ($\sim$6 meV).

In common with other layered superconductors, the thermodynamic properties of the cuprates 
are strongly anisotropic with respect to the direction of the applied magnetic field \cite{junod2}.  
A field perpendicular to the conducting planes generally causes a stronger suppression 
of the superconductivity than one applied within the plane. To see whether the resonance 
intensity depends similarly on field direction, we imposed a 6.8 T field parallel to the $ab$-plane (Fig. 1f).  
A noticeable, but much smaller suppression ($\leq 10\%$) of the resonance is observed. Thus, the anisotropy of the field effect on the resonance reflects the anisotropy of other physical superconducting properties.   This is the clearest evidence to date that the resonance is very sensitive to the superconducting pairing. 

Although the constant-$Q$ difference spectra (Fig. 1) are excellent 
for determining the intensity gain of the resonance from the normal to 
the superconducting state, such measurements do not provide information 
about how the field affects the momentum distribution of the excitations. 
We therefore scanned wave vector in the $B\sim ||c$-axis geometry (Fig. 1a) 
at energy transfers below, at, and above the resonance. At 150 K, the 
constant-energy scans at the resonance energy (Fig. 2a) show a broad peak 
centered at $(\pi,\pi)$ on a sloped linear background. There are no observable 
intensity differences in the profiles for $B = 0$ and 6.8 T. On cooling to 70 K 
(Fig. 2b), the resonance narrows in width and grows in intensity, but is also 
clearly suppressed at 6.8 T. At 10 K (Fig. 2c), the suppression of the resonance 
at 6.8 T is more severe. Direct inspection and fits of Gaussian profiles to the data (Fig. 2, a-c) indicate no appreciable field-induced change in the momentum dependence of the resonance peak. In contrast, the thermal fluctuations associated with warming from 10 K to 70 K yield a decrease in $Q$-integrated intensity which is numerically similar to that found on application of a 6.8 T field at 10 K, and 
makes the profile considerably broader.

Below the 20 meV spin gap in the superconducting state \cite{dais}, 
the constant-energy scan at 16 meV in the 6.8 T (Fig. 2d) is featureless, 
that is, there are no strong field-induced AF correlations at this energy. 
At 24 meV, where incommensurate spin fluctuations were observed \cite{dai2,mook1} 
the field seems to suppress the constant-energy profiles in both the normal (Fig.  2e) and 
the superconducting state (Fig. 2f). The effect is more difficult to see than at the 
resonance energy although it could represent a similar fractional change in a 
considerably smaller signal above background. Finally, when we move to an energy 
just above the resonance, we discern no field-induced change in the $Q$-dependent 
profiles at both 70 K and 10 K (Fig. 2g and h).

To demonstrate the effect of a magnetic field on the energy dependence of the excitations 
in (123)O$_{6.6}$, we show the experimentally measured difference spectra, 
$S(Q_0,\omega,B = 0 {\rm T}) - S(Q_0,\omega,B = 6.8 {\rm T})$, 
for $B\sim ||c$-axis (Fig. 3, a-c) and in the ab-plane (Fig. 3d and e) at different temperatures. 
Since one expects the lattice contribution (phonons) to the scattering to be field-insensitive, 
this procedure will yield the net change of the magnetic intensity induced by the field. At 150 K, a 
6.8 T field has no effect on the spin fluctuations in the measured energy range for $B\sim ||c$-axis (Fig. 3a). 
On cooling to 70 K, the difference spectra show a broad peak for $B\sim ||c$-axis (Fig. 3b) and no 
discernible feature for $B||ab$-plane (Fig. 3d). Higher resolution measurements in the $B\sim ||c$-axis 
geometry at 6.2 T field (Fig. 3f) confirm the result of Fig. 3b. In the superconducting state 
(Fig. 3c, e, and g), the difference spectra show a sharp peak at 34 meV for both field orientations, again confirming that the resonance is the feature most obviously affected by the magnetic field for spin fluctuations 
below 50 meV. In addition, the sharpness here contrasts with the diffuse nature of the difference just above $T_c$, thus showing the sensitivity of the resonance width, or decay rate, to true bulk superconductivity even in a material with an extended pseudogap regime above $T_c$. Simple summation as well as least-squares fits to the data 
(Fig. 3f and g) indicate that the energy-integrated effect of the magnetic field just above $T_c$ is 
comparable to that well below $T_c$. The narrowing of the widths in the field-induced 
difference spectra together with the approximate preservation of weight in these spectra 
suggest that the resonance width depends strongly on the phase 
coherence time, while its integrated weight measures the pair density. 

In Fig. 4 we plot the temperature dependence of the neutron scattering 
intensity at the resonance energy at zero and 6.8 T field. Most growth in the 
field-dependent peak signal tracks the growth in the resonance itself on entering the 
superconducting state. The inset shows the field dependence of the normalized intensity. 
The solid line is a linear fit to the data, $i.e.$, $I/I_0 = 1-B/B_{char}$ with $B_{char} = 36$ T. 
Because $B_{char}$ is not far from the upper critical field $B_{c2} = 45$ T estimated for our sample, 
it is reasonable to associate the sharp resonance peak with the superconducting volume 
fraction which for simple superconductors is proportional to $1-B/B_{c2}$. 
 
Figures 2 and 3 indicate that wherever we look at fixed temperature, 
the external field either has no effect or reduces the magnetic scattering signal.  
The total moment sum rule \cite{hayden,aeppli} states that the magnetic structure factor, 
when integrated over all wave vectors and energies should be a temperature- and field-independent constant. 
Therefore, compensatory increases in signal should occur elsewhere in $Q$-$\omega$ space. 
Possible destinations for the lost spectral weight include elastic scattering, new or 
softened optic modes, a subgap continuum, and other continuum scattering spread sufficiently over $Q$ 
and $\omega$ space so as to be unobservable at any particular $Q$ and $\omega$. 
Searches for the most obvious elastic scattering, namely that due to the proposed field-induced AF order in cuprate superconductors \cite{arovas}, have been fruitless. 
Scans in the $[H, K, 0]$ plane with $B||c$-axis revealed no evidence for optic mode scattering 
at $\hbar\omega  = 34$, 41, and 50 meV apart from that already present at zero field. In addition, we have not yet found field-induced subgap scattering (Fig. 2d). While our searches for the missing spectral weight were by no means exhaustive, a 6.8 T field does not induce obvious new excitations around $(\pi,\pi)$ with energies between 10 and 50 meV that are big enough to compensate the spectral weight loss of the resonance. It is noteworthy 
that similar sum rule violations also exist for spectra collected using angle resolved photoemission \cite{anderson1}, scanning tunneling microscopy \cite{renner}, and infrared spectroscopy \cite{bosov} in some high-$T_c$ superconductors. However, we caution that an explanation of the superficially similar effects in these measurements must take into consideration that different experimental technique probe quite different correlations. 

We have discovered that a very modest field applied to (123)O$_{6.6}$ 
yields a very significant reduction in the 34 meV resonance. Therefore, the 
resonance can be reduced either by warming or applying a magnetic field, with the highest 
amplitude and narrowest profiles both in energy and wave vector occurring in the bulk superconducting state. 
Even more surprising is the persistence of the effect above $T_c$ (Figs. 2-4). 
Although there are no explicit microscopic predictions about the effects of magnetic fields on the resonance, our finding has important implications for understanding the mechanism of high-$T_c$ superconductivity.  
In particular, the larger $c$-axis field effect presents another challenge to the interlayer tunneling theory, according to which the resonance should be strongly affected by the in-plane field that 
disrupts the coherent Josephson coupling along $c$-axis \cite{yin}. 
On the other hand, the data are consistent with mechanisms where the dominant loss of entropy on entering the superconducting state is due to the growth of magnetic correlations in the CuO$_2$ planes. 
A modest field reduces the resonance just as it reduces the specific heat anomaly \cite{junod1} near $T_c$, 
so that the temperature derivative of the resonance intensity tracks the specific heat not 
only as the hole density is varied \cite{dais}, but also as the magnetic field is imposed 
for fixed $x$ (ref. \cite{janko}). The anisotropic field effect 
reflects the anisotropy of other physical properties such as the magnetic penetration depth and coherence length. Thus, the persistence of the effect above $T_c$ would naturally be due to 
superconducting fluctuations or incoherent pairing which also display this 
anisotropy in the normal state. Such fluctuations exist in models postulating a Bose-Einstein condensation of pre-formed pairs at Tc in the underdoped cuprates \cite{uemura,emery1,lee}.

{\noindent \bf Acknowledgements}

{\noindent We appreciate discussions with B. Janko, A. Junod, A. Kapitulnik, M. V. Klein, Ch. Renner, and D. J. Scalapino. The work at ORNL was supported by the US DOE.}

{\noindent
Correspondence and requests for materials should be addressed to P. Dai. (e-mail: piq@ornl.gov).
}

\begin{figure}
\caption{Reciprocal lattice diagram and neutron scattering results. For the experiment, 
we used the HB-3 triple-axis spectrometer at the High-Flux Isotope Reactor of Oak Ridge National Laboratory. 
The collimations were, proceeding from the reactor to the detector, 48-40-60-120 min 
(full-width at half-maximum) and the final neutron energy was fixed at either $E_f  = 30.5$ meV or 14.78 meV. 
The monochromator, analyzer and filters were all pyrolytic graphite. The low 
temperature data were collected after field cooling the sample from 200 K. {\bf a}, 
To determine the effect of a field along the $c$-axis on the acoustic spin fluctuations, we first 
align the crystal in the $[H, K, 0]$ zone. 
The sample
is then rotated $\alpha$ degrees around the $[3H,-H,0]$ direction,
where $\alpha=\arctan[(2\pi L/c)/(2\pi\sqrt{H^2+(3H)^2}/a)]\approx 20.6^\circ$ 
for $H=0.5$ and $L=1.8$ rlu.
 such that  the horizontal scattering plane is spanned by wave vectors along the 
$[1,3,\tan\alpha\sqrt{10}  c/a]$ and $[3,-1,0]$ directions. In this geometry, we can perform both 
constant-$Q$ (Fig. 1b and c) and constant-energy (Figs. 2, a-h) scans at the acoustic mode intensity maximum 
($L = 1.8$ rlu) while still keeping 94\% of the applied (vertical) field along the $c$-axis ($B\sim ||c$-axis). 
To see the influence of a field along $c$ on the intensity of the optical mode, we oriented the $c$-axis 
of the crystal perpendicular to the scattering plane ($B||c$-axis). This geometry allows 
any wave vector of form $(H, K, 0)$, that is, at the $L = 0$ intensity maximum of the optical mode, to be reached. 
{\bf d}, Conventional $[H, 3H, L]$ zone geometry where the applied field is in the ab-plane along the $[3, -1, 0]$   direction. {\bf b}, {\bf c}, Difference spectra of the neutron intensities between $T = 10$ K ($< T_c$) 
and $T = 70$  K ($T_c + 7.3$ K) at wave vector $Q = (0.5, 1.5, 1.8)$ for zero and 6.8 T field in the 
$B\sim ||c$-axis geometry.  {\bf e}, {\bf f}, Similar data in the $B||ab$-plane geometry. Because the differences in sample geometries, the actual intensity gain of the resonance in the $B||ab$-plane experiment is slightly 
higher than that of the $B\sim ||c$-axis case at zero field for the same incident beam monitor (300 counts which correspond to $\sim$8 min per point at $\hbar\omega  = 34$ meV). We normalized the intensity gain of the resonance in these two experiments at zero field and multiplied the same scale factor to the finite field case.
 }
\label{autonum}
\end{figure}

\begin{figure}
\caption{Effect of a magnetic field on momentum dependence of the spin excitations. Constant-energy scans of the neutron scattering intensity as a function of wave vector in the $B\sim ||c$-axis geometry. {\bf a-c} 
show the results at the excitation energy of 34 meV; {\bf d}, 16 meV; {\bf e}, {\bf f}, 24 meV; and {\bf g}, 
{\bf h}, 41 meV. The open and filled circles represent data for identical scans at zero and 6.8 T field, respectively. The dashed and solid lines are Gaussian fits to the data.
}
\end{figure}

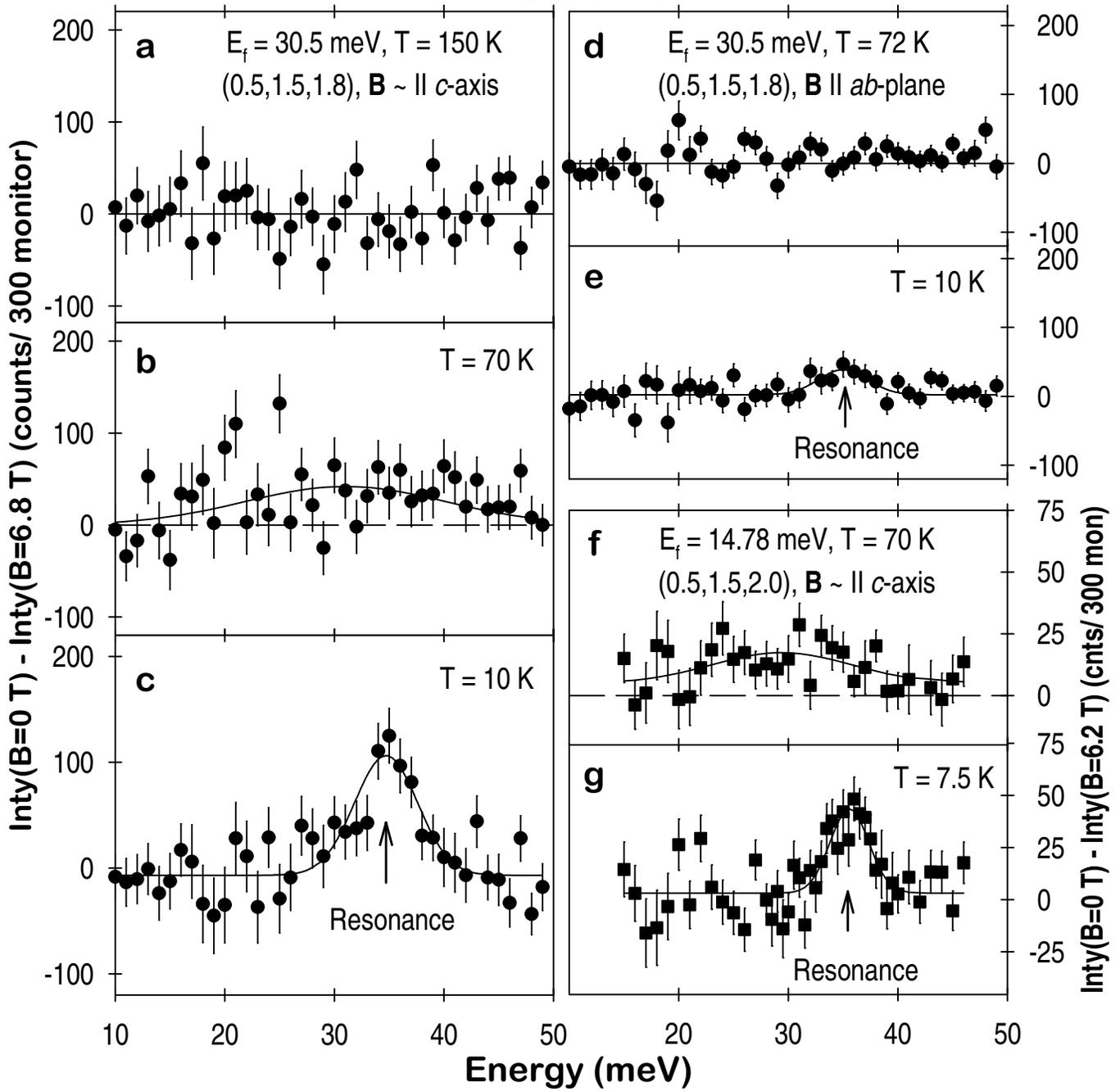
\begin{figure}
\caption{Effect of a magnetic field on the energy dependence of the 
spin excitations. Difference spectra of the neutron scattering intensities 
between zero and finite magnetic field in {\bf a-}c and {\bf d-e} 
show the low-resolution results obtained with $E_f  = 30.5$ meV at wave vector 
$Q = (0.5, 1.5, 1.8)$ in the $B\sim ||c$-axis and $B||ab$-plane geometries, respectively. 
High-resolution measurements in the $B\sim ||c$-axis geometry with $E_f  = 14.78$ meV are 
shown in {\bf f-g}. The scattering should be centered around zero as shown by the solid lines 
in {\bf a} and {\bf d} if spin fluctuations are not affected by the field. The solid lines in {\bf c} and 
{\bf e-g} are unconstrained least-squares fits to the data by Gaussians on floating constant backgrounds. 
The solid line in {\bf b} is a guide to the eye. Positive scattering (above zero) indicates a net loss in the acoustic magnetic spectral weight when the field is applied. Strong phonon scattering at $\sim$20 meV 
in the raw constant-$Q$ scans yielded the large error bars in the difference spectra around that energy.
}
\end{figure}

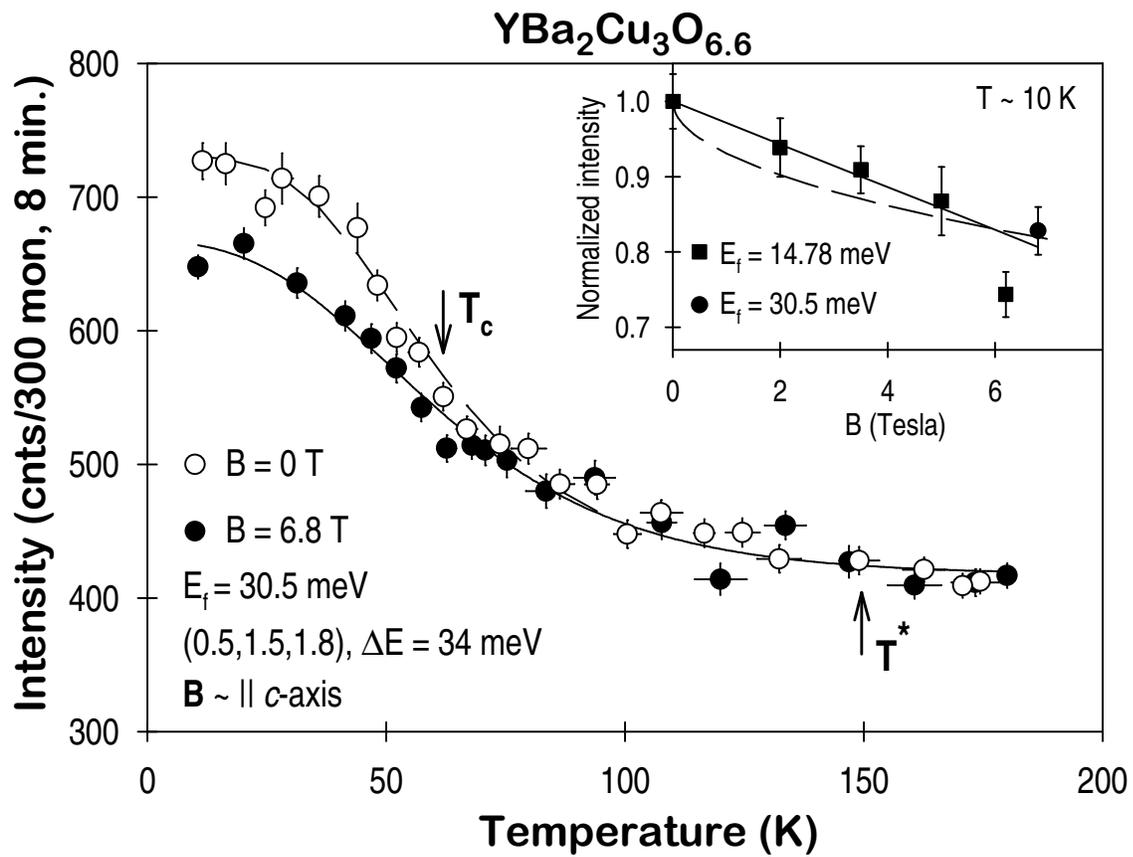
\begin{figure}
\caption{
Effect of a magnetic field on the temperature dependence of the resonance 
and the field dependence of the resonance at $\sim$10 K. The data are collected in the $B\sim ||c$-axis geometry.  
The open and filled circles show results at $Q = (0.5, 1.5, 1.8)$ and $\hbar\omega  = 34$ meV for 
zero and 6.8 T field, respectively. The pseudogap temperature $T^\ast$ and $T_c$ of the 
sample are indicated by arrows. The solid and dashed lines are guides to the eye. The inset shows the 
field dependence of the normalized resonance intensity at approximately 10 K. The filled squares are high-resolution data at $Q = (0.5,1.5,2)$ with $E_f  = 14.78$ meV. The solid line corresponds to $I/I_0 = 1-(B/B_{char})$ 
with $B_{char} = 36$ T while the dashed line represents $I/I_0 = 1-(B/B_{char})^{1/2}$, where $B_{char} = 208$ T. 
}
\end{figure}

\end{document}